\documentclass{aa}  

\usepackage{float}
\usepackage{graphicx}
\usepackage{txfonts}
\usepackage{natbib}
\bibpunct{(}{)}{;}{a}{}{,}

\usepackage{booktabs}
\usepackage{makecell}
\usepackage{color, colortbl}

\begin{document}

   \title{Ponderomotive forces in magnetized non-thermal space plasmas due to cyclotron waves }


   \author{Joaquín Espinoza-Troni
          \inst{1}
          \and
          Felipe A Asenjo\inst{2} \and Pablo S Moya \inst{1} \fnmsep}
          

\institute{Departmento de Física, Facultad de Ciencias,
Universidad de Chile, Santiago, Chile\\
\email{joaquin.espinoza.t@ug.uchile.cl; pablo.moya@uchile.cl}
\and
Facultad de Ingenier\'ia y Ciencias,
Universidad Adolfo Ib\'a\~nez, Santiago, Chile. \\
\email{felipe.asenjo@uai.cl}
}

\date{}

 
  \abstract   {The ponderomotive force is involved in a variety of space plasmas phenomena which are characterized by the family of Kappa distributions. Therefore, evaluating these nonthermal effects in the ponderomotive force is required.}
{The Washimi and Karpman ponderomotive interaction due to cyclotron waves is evaluated for different space conditions
considering low-temperature magnetized plasmas described by an isotropic Kappa distribution and with a wave propagation parallel to the background magnetic field.}
{We performed a brief analysis of the influence of the Kappa distribution in the dispersion relation for a low-temperature plasma expansion at the lowest order in which the thermal effects are appreciated without considering the damping characteristics of the wave. The different factors of the ponderomotive force are obtained and analyzed separately as a function of the wavenumber, the spectral index $\kappa$, and the plasma beta.}
{We have found a relevant influence of the non-thermal effects in all the factors of the ponderomotive force for magnetized plasmas. The effect of the kappa distribution has been evaluated for a wide variety of space environments as the solar wind and the different regions of our magnetosphere where it has been found that these results can be relevant for the solar wind, the magnetosheath, the plasmasheet, and the polar cusps. We have also analyzed the role of the non-thermal effect in the induced Washimi and Karpman ponderomotive magnetization in the context of spatial plasmas and the total power radiated associated with it.}
{We find that even for nearly cold magnetized plasmas and waves far from the resonances the effect of the kappa
parameter in the ponderomotive force cannot be neglected. This suggests a significant role of the Kappa distribution in ponderomotive phenomena of space physics.}


   \maketitle
%

\section{Introduction}

Space plasma environments are constantly affected by wave-wave and wave-particle interactions. Indeed, the solar wind plasma and its interplanetary magnetic field are externally forcing our planetary magnetosphere by the excitation of waves and the penetration of particles that propagate, for example through the magnetic field lines of the polar cusps or the nightside magnetotail \citep{Lakhina_1990,Chen_1992}, or by instabilities in the solar wind and the geomagnetic field that leads to the propagation of Ultra Low Frequency waves (ULF) in our magnetosphere \citep{Hughes_2013,DiMatteo_Villante_Viall_Kepko_Wallace_2022}. Usually, these processes are studied for low amplitude perturbations to simplify the analysis of the dynamics by treatment of linearized systems. Nevertheless, this approach does not take account of a variety of phenomena that occur in space plasmas when we consider finite amplitude waves. Thus, when nonlinear terms are considered, a diversity of effects in space plasmas emerge, such as three-wave decay interactions, modulations instabilities, self-wave interactions, wave trapping, and density cavities, among others \citep[see][]{Wong_1982,Eliasson_Shukla_2006,Kamide_Chian}.

Also, when non-linear effects are relevant, we need a useful tool that allows us to describe mathematically the space physics dynamics between waves and particles. To solve this and to take account of the nonlinear perturbative terms of the electromagnetic fields that emerge in the Lorentz force the concept of ponderomotive force has been developed and widely used \citep{kentwell_time-dependent_1987}. Ponderomotive forces are time-averaged non-linear forces that emerge from the interaction of quasi-monochromatic waves or spatially inhomogeneous monochromatic waves with plasma. They are a useful tool that allows us to study the complex dynamics due to the interaction of waves with plasma in a slow time scale concerning the carrier high frequency of the wave, which can simplify our understanding of some physical phenomena \citep{lundin_ponderomotive_2007}. Due to this, the ponderomotive force has had great relevance in the study of phenomena that involve the interaction between waves and plasma in different environments.  It has been widely used in lasers, where it causes self-focusing effects \citep{karpman_two-dimensional_1977,washimi_method_1989,rezapour_zahed_mokhtary_2018,gupta_kumar_bhardwaj_2022}, in addition to having applications in phenomena such as laser ignition of controlled nuclear fusion \citep{hora_2007,hora_miley_lalousis_moustaizis_clayton_jonas_2014,hora_2016}, among others.

The ponderomotive force has also significant applications in space physics and is responsible for a diversity of phenomena \citep[see][]{lundin_ponderomotive_2007,lundin2022cosmic}. It has been used to study the redistribution of plasma in the terrestrial magnetosphere and the acceleration of ions in the polar wind  \citep{Allan1992,Li_Temerin_1993,guglielmi_ponderomotive_2001,lundin_ponderomotive_2007,nekrasov_nonlinear_2012,nekrasov_nonlinear_2014,Guglielmi_Feygin_2023}. Some of these theoretical models are inspired by measurements of the satellites Freja and Viking in the ionosphere \citep{lundin_hultqvist_1989,lundin_gustafsson_eriksson_marklund_1990}; however, measurements are lacking in the Earth's magnetosphere that confirms experimentally the main points exposed in these works, for which they have been proposed some indirect measurement methods based on the dependence of the foreshock locations on the orientation of the field lines of the interplanetary magnetic field \citep[see][]{guglielmi_impact_2018}. In addition to the magnetosphere, the ponderomotive forces have also been studied on the Sun, where they are relevant to understanding the difference in ion composition between the photosphere and the solar corona \citep{laming_2004,laming_2015}. They have also been suggested to explain the Magnetic Holes (MHs) and Magnetic Decreases (MDs) of the interplanetary magnetic field observed in the solar wind due to the interaction of the plasma with phase steepened Alfvén waves \citep{Tsurutani_Dasgupta,dasgupta_2003}. They are involved in the nonlinear coupling of large amplitude electromagnetic pump waves with low-frequency collisional modes in the ionosphere \citep{Drake_1974,Stenflo_1990}. Also, they have been utilized to explain the magnetic-field-aligned electron density compressions by dispersive shear Alfvén waves (DSAWs) observed by FREJA and FAST spacecraft in the magnetosphere \citep{Stasiewicz_Bellan_Chaston_Kletzing_Lysak_Maggs_Pokhotelov_Seyler_Shukla_Stenflo_etal._2000,Shukla_Stenflo_Bingham_Eliasson_2004}. The ponderomotive force interaction with plasmas can also act as a generator of slowly varying magnetic fields \citep{washimi_magnetic_1977} which had been widely studied because of its importance in the magnetic field generation in laser-matter interaction and dense plasmas in astrophysical compact objects \citep{na_temperature_2009,shukla_generation_2010,jamil_karpman-washimi_2019}.

Because the ponderomotive force is generated due to the electromagnetic perturbation of the plasma, its characteristics depend as much on the wave modulation and on the interaction with the medium described by the dielectric tensor and its dispersion relation. At the same time, these macroscopic variables ultimately depend on the velocity distribution that characterizes the plasma. It is well known that in most space environments, because the plasma does not reach the thermal equilibrium due to its low collision rate, instead of being characterized by the Maxwellian distribution it is described by the family of Kappa distributions \citep{Viñas_Mace_Benson_2005,Nieves‐Chinchilla_Viñas_2008,Yoon_2014,Espinoza_2018,lazar_kappa_2021,Eyelade_2021,Lazar_López_Poedts_Shaaban_2023}. This family of distributions depends on the $\kappa$ parameter and can be considered as a generalization of the Maxwellian distribution that it is recovered when we tend $\kappa$ to infinity. They have been widely observed experimentally in near-Earth space environments, in addition to being proposed as an explanation of a great variety of phenomena such as the heating of the corona by velocity filtration and the acceleration of the fast solar wind \citep{pierrard_lazar_2010}. In many space environments the kappa parameter can have low values in the interval of 2-7 \citep{Livadiotis_2015}, even in the inner magnetosphere around the plasmapause the kappa parameter can have values of $\kappa \sim 10$ \citep{Kirpichev_Antonova}.  Its theoretical support is still a reason for discussion in the scientific community, and some models have been developed in an attempt to explain this observation. Among them, an attempt has been made to extend statistical mechanics, with what is known as superstatistics \citep{Davis_Avaria_Bora_Jain_Moreno_Pavez_Soto_2019,Gravanis_Akylas_Livadiotis_2020,Yoon2021}. Therefore, due to its great importance in the description of collisionless plasmas in space environments, its effect has to be evaluated when studying certain phenomena in space physics. Moreover, the non-thermal effects of the plasma described by the Kappa distribution can significantly affect the behavior of the ponderomotive force and thus the phenomena associated with it. Therefore an investigation of the effect of the Kappa distribution in the ponderomotive force is required.

Indeed, recently a detailed analysis of the characteristics of ponderomotive forces in non-thermal unmagnetized plasmas has been developed \citep[see][]{Espinoza-Troni_A_Asenjo_Moya_2023}. A first approach has been made to the study of the non-thermal effects in the ponderomotive force by including the contribution of the Kappa distribution on non-magnetized plasmas considering the movement of electrons in a background of immobile ions. It demonstrated the importance of the kappa parameter in the interaction between plasmas and waves of inhomogeneous amplitude in time and the magnitude of the induced magnetic fields derived from this interaction. Also, those results show that for unmagnetized plasmas, the non-thermal effects are negligible for the spatial ponderomotive force when non-relativistic thermal velocities are considered. Nevertheless, it is expected that including other parameters that characterize the interaction of the plasma with the waves,  different behavior of the effect of the kappa parameter in the ponderomotive force could be obtained. Indeed, that work can be extended to study the case of magnetized plasmas. This case becomes more relevant in space physics where the plasma is usually interacting with an external magnetic field. The purpose of this work is to give a detailed analysis of the effect of including the Kappa distribution in the ponderomotive force due to the interaction of magnetized plasmas with waves propagating parallel to the external magnetic field and to evaluate its implications in different space environments. We include the non-thermal effects in the ponderomotive force by expanding the dielectric tensor for Kappa distributions in low-temperature magnetized plasmas without considering the damping of the wave. Then, we compare the magnitude of the factors accompanying the different terms of the ponderomotive force for the Kappa and Maxwellian distribution, for different values of the plasma beta and the ratio between the plasma frequency and gyrofrecuency to model the different space conditions. 

This article is organized as follows: Section 2, includes the non-thermal effects in the kinetic dielectric tensor, and we obtain asymptotic expressions for low temperatures. We also give a brief analysis of the influence of the kappa parameter in the solutions of the dispersion relations of waves propagating parallel to the external magnetic field. Then, in Section 3 we include the dielectric tensor in terms of the ponderomotive force and we deduce its expressions. Later we give a general analysis of the influence of the non-thermal effects, for each term of the ponderomotive force and wave mode, and compare our results with the thermal (Maxwellian) case. Also, we use this analysis to study the nonlinear perturbation of the magnetic field by the ponderomotive force produced by electromagnetic waves propagating parallel to the external magnetic field. In section 4 we evaluate our results for different space environments for its typical plasma parameters. Finally, in Section 5 we summarize the main conclusions of this work.

\section{Dispersion relation for thermal and non-thermal plasmas with low temperature}

We consider a Kappa distribution for isotropic three-dimensional plasmas \citep[][among others]{hellberg_generalized_2002,Yoon2006,hau_fu_chuang_2009,Yoon2012,vinas2015,Lazar2016,Moya_2021}
\begin{equation}
    f_{\kappa s}(\mathbf{v}) = \frac{n_s}{\pi^{3/2}\alpha_s^3 \kappa^{3/2}}\frac{\Gamma(\kappa+1)}{\Gamma(\kappa-1/2)}\left(1 + \frac{v^2}{\kappa \alpha_s^2}\right)^{-(\kappa + 1)}\, .
\label{eq:DistribucionKappa}
\end{equation}
Here, $f_{\kappa s}$ is the Kappa distribution for the species $s$, $n_s$ is their number density, $\alpha_s\, \{= \sqrt{2k_B T_s /m_s}\}$ is their thermal velocity, $k_B$ is the Boltzmann constant, $m_s$ is their mass, $T_s$ is their temperature and $\Gamma$ is the Gamma function. 
    
In this Section, we give a brief review of the dispersion relation for high-frequency waves propagating through low-temperature plasmas described by a Kappa distribution (\ref{eq:DistribucionKappa}) and with a background magnetic field parallel to the wave propagation. As a first case, we consider only the dynamics of electrons with a static background of ions to achieve quasi-neutrality. Thus, we neglect the contribution of the ions in the dispersion relation i.e. we are considering frequencies much larger than the ion plasma frequency. Due to the huge mass of the ions in comparison to the electrons, this consideration is valid for many contexts, nevertheless in some cases in which the ponderomotive force is involved as in the ULF waves the contribution of the ions can not be neglected \citep{Guglielmi_Hayashi_Lundin_Potapov_1999,nekrasov_nonlinear_2012,guglielmi_impact_2018,Guglielmi_Feygin_2023}. For this last case when the wave frequency is comparable to the ion plasma frequency we obtain the ion cyclotron wave mode of propagation. Therefore we are analyzing the effect of the Kappa distribution for electron modes and ion cyclotron waves for a parallel propagation with the background magnetic field \citep{chen_1984}, which will be useful as a first approximation in the study of the non-thermal effect in the ponderomotive force of magnetized plasmas. Also, to delimit our investigation in this work we focus on analyzing the right-handed modes of propagation for the electron waves and the left-handed mode of propagation for the ion cyclotron waves. A detailed deduction of the dielectric tensor and the dispersion relation for non-relativistic magnetized plasma modeled by isotropic Kappa distributions can be found in \citet{mace_dielectric_1996}. The dispersion relation for Kappa distributed plasmas has also been investigated in a variety of different contexts \citep{hellberg_mace_baluku_kourakis_saini_2009,pierrard_lazar_2010,kourakis_sultana_hellberg_2012,lazar_kourakis_poedts_fichtner_2018}. The dielectric tensor for magnetized plasmas can be deduced using kinetic theory by linearly perturbing the Vlasov equation. In this way, the same result as \citet{hellberg_generalized_2002} is obtained. Namely
    \begin{equation}
    \varepsilon_\pm(k,\omega) = 1 + \sum_s \frac{\omega_{ps}^2}{\omega k \alpha_s}Z_{\kappa M}\left(\frac{\omega \pm \Omega_s}{k\alpha_s}\right)\, ,
    \label{eq:varepsilon}
\end{equation}
where the upper sign is associated with the right-handed waves and the lower for the left-handed waves.  Here,  $\omega$ is the frequency, $k$ is the wave number, $c$ is the velocity of light,  $Z_{\kappa M}$ is the modified generalized plasma dispersion function \citep{hellberg_generalized_2002}, $\omega_{ps}$ and $\Omega_s$ are the plasma frequency and the gyrofrecuency for the species $s$ respectively. We have used $\varepsilon_\pm = \varepsilon_{11} \pm i \varepsilon_{12}$ as the eigenvalues of the dielectric tensor for the transversal modes of propagation, where $\varepsilon_{ij}$ are the components of the dielectric tensor with a magnetic field in the $z$ direction. The dispersion relation is given by $\varepsilon_\pm = k^2c^2/\omega^2$. Because the longitudinal waves do not interact with the magnetic field they remain the same as the unmagnetized case analyzed for the ponderomotive force in \citet{Espinoza-Troni_A_Asenjo_Moya_2023}. \\ 

We consider low-temperature plasmas such that  $(\omega \pm \Omega_s)/k\alpha_s \gg 1$ for both ion and electron species. Under this approximation, we can make use of the asymptotic expansion of the generalized plasma dispersion function for large arguments (see appendix A), and truncate the series at the lowest order in which the effect of the kappa parameter appears, so that the temperature is included in the dielectric tensor. In this way, we expand in $k\alpha_s/(\omega \pm |\Omega_s|)$ terms in second order in the dielectric tensor.  As a first approach to the study of the
non-thermal effects on the ponderomotive force for magnetized plasmas, we do not consider the imaginary terms
i.e. the damping characteristics of the wave in the expansion of the generalized plasma dispersion function which we will leave for future research. We can neglect the damping characteristics of the wave as long as we consider $k^{-1}(\omega-|\Omega_s|) \gg \alpha_s$ \citep{fitzpatrick_2015}; which, considering that due to our low-temperature approximation, we have low values of $\alpha_s$, this is satisfied if the frequency is far from the resonances. 
    
By considering the electron species and only one type of ion, we have the approximated dielectric tensor components for electromagnetic waves in magnetized plasmas with parallel propagation \citep{chen_1984} with finite and low temperature, and for right-handed (upper sign) and left-handed (lower sign) waves, as follows $\varepsilon_{\pm}(\omega,k) = \varepsilon_{0\pm}(\omega) - ({k^2c^2}/{\omega^2})\delta_{\pm}(\omega)$. Where $\varepsilon_{0\pm}$ is the dielectric component for cold plasmas and the $\delta(\omega)$  factor is where the effect of the finite temperature is contained and is responsible for the spatial dispersion of the wave. Considering that the dispersion relation is given by $\varepsilon_{\pm} = k^2 c^2/\omega^2$ we can directly express the dielectric component as a function of the frequency. Also, due to our low-temperature plasmas approximation far from resonances where we have to consider that $(\omega-\Omega_s)/k\alpha_s \gg 1$, then  $\delta_\pm \ll 1$ for both electron waves and ion cyclotron waves. Hence, for the plasmas that we are considering we can use as a good approximation
\begin{equation}
    \varepsilon_\pm \approx \varepsilon_{0\pm}(1-\delta_{\pm}).
    \label{eq:varepsilon_approx}
\end{equation}

Next, we will give a review of the main characteristics of the modes of propagation that we will use in this work and we will limit the range of frequencies for which our approach is valid.

\subsection{Electron waves}

As we said above, we can neglect ions as long as the ion-plasma frequency is very low compared to the wave frequency $\omega_{pi} \ll \omega$  \citep{krall_trivelpiece_1986}. Using this approximation and considering the right-handed wave mode of propagation we can obtain that
\begin{equation}
    \varepsilon_{0} = 1 - \frac{(\omega_{pe}^2/\Omega_e^2)}{y(y-1)}, \qquad \delta = \frac{\beta_e}{2} \left(\frac{\kappa}{\kappa-3/2}\right)\frac{y}{(y-1)^3},
    \label{eq:varepsilonColdEC}
\end{equation}
where  $y = \omega/|\Omega_e|$ and $\beta_e = 8\pi n_e k_B T_e/B_0^2$ is the electron plasma beta, with $B_0$ the norm of the background magnetic field. We normalize the wavenumber as $x = kc/\omega_{pe}$. We can notice that when we set $\beta_e = 0$, i.e. when we do not consider the finite temperature effects, we recover the usual dispersion relation for cold magnetized plasmas with parallel wave propagation \citep{bittencourt_2010}. \\
\begin{figure*}
\centering
\includegraphics[width=\linewidth]{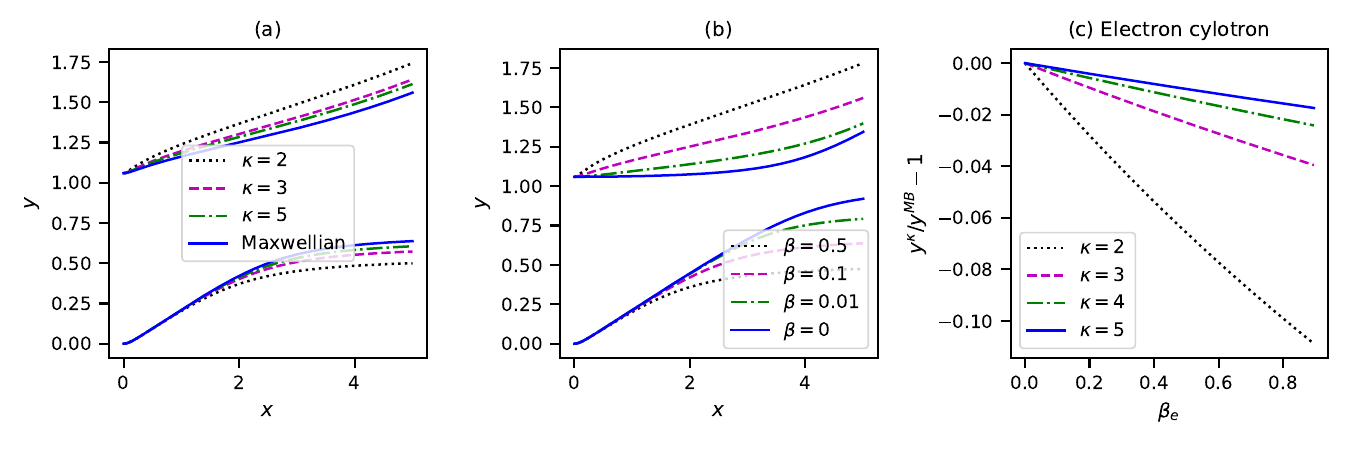}
    \caption{Solution for the dispersion relation for Right handed electron waves for $y$ as a function of $x$ with $\omega_{pe}/|\Omega_e| = 0.25$, (a) for different values of $\kappa$ with $\beta = 0.1$,  (b) for different values of $\beta$ with $\kappa \rightarrow \infty$.  (c) The relative difference of the Kappa and Maxwellian scaled frequencies $y^\kappa/y^{MB}-1$ for right-handed electron waves as a function of $\beta_e$ with $\omega_{pe}/|\Omega_e| = 30$ and $x = 0.3$ for the electron cyclotron solution. }
\label{fig:RD}
\end{figure*}

We can solve numerically the dispersion relation and we get two solutions for the right-handed mode of propagation as we expected due to the cold case which we can see in figure \ref{fig:RD} which shows the solutions of the scaled frequency $\omega/|\Omega_e|$ as a function of the scaled wavenumber $kc/\omega_{pe}$. The lower branch corresponds to the electron cyclotron waves, and the Upper branch is associated with the solution for the unmagnetized case, which is recovered when we consider $|\Omega_e| = 0$. Figure \ref{fig:RD}.(a) gives the solutions of the dispersion relation for different values of kappa and the figure \ref{fig:RD}.(b)  for different values of the beta parameter. We can notice in these figures that the waves cannot propagate in certain bands of frequencies between both branches, also this interval increases with the temperature and the non-thermal effect. Also, we can notice that for $\beta_e = 0$ we recover the resonance point at $\omega = |\Omega_e|$ for cold plasmas. This figure was made for underdense plasmas for values of $\omega_{pe}/|\Omega_e| \sim 0.25$, for illustrative purposes, because in this regime these characteristics are more notorious in the plot. Nevertheless, in most space environments we found overdense plasmas with $\omega_{pe}/|\Omega_e| > 1$ (see below).  \\

We are going to clarify the range of the parameter values that we are taking into consideration in this work. Notice that our condition for the asymptotic expansion of the modified generalized plasma dispersion function is equivalent to $\beta_e^{1/2}x/(1-y) \ll 1$. Therefore, due to the dispersion relation, the validity of our approximation will depend on the values of the wavenumber, the plasma beta, $\omega_{pe}/|\Omega_{e}|$, and the mode of propagation. For the electron cyclotron solution the dependence of $\beta_e^{1/2}x/(1-y)$ in the ratio of frequencies is negligible. Also, we have neglected fourth order terms of $\zeta^{-1} = \beta^{1/2}x/(1-y)$ in equation \ref{eq:varepsilon}. Hence, for our approximation to be consistent we are going to consider $\zeta^{-4}<0.01$. In table \ref{tab:RangeOfValidity} we have for different values of the wavenumber the maximum beta value that accomplishes this condition for the electron cyclotron waves. We can see that for this mode of propagation, our results will be valid for larger values of the plasma beta as long as we consider lower values of the frequency. Also, in this work, we will focus mainly on analyzing overdense plasmas with $\omega/|\Omega_e| > 10$, which are typical values in space environments as we would see in section 4. For that case, we have that $|\varepsilon_0| \gg 1$ for electron cyclotron waves, so we can use as a good approximation that $\varepsilon_0 \approx - (\omega_{pe}^2/\Omega_e^2)/y(y-1)$.  \\

\renewcommand{\tabcolsep}{20pt}

\begin{table}
\centering
\caption{Maximum value of the electron plasma beta $\beta_{\text{max}}$ that accomplish the condition $\zeta^{-4} < 0.01$ for different values of the scaled wavenumber $x$ and for electron waves, where $\zeta^{-1} = \beta^{1/2}x/(1-y)$. In the third column, we have also their respective value of $y$ for $x$ and $\beta_{\text{max}}$ with $\omega_{pe}/|\Omega_e| = 30$.}
\label{tab:RangeOfValidity}
\begin{tabular}{ccc}
\toprule
$x$ & $\beta_{\text{max}}$ & $y$ \\ 
\midrule
0.2 & 2.3 & 0.04 \\
0.3 & 0.9 & 0.08 \\
0.35 & 0.65 & 0.10 \\
0.4 & 0.45 & 0.13 \\
0.5 & 0.25 & 0.19 \\
0.6 & 0.15 & 0.26 \\
0.7 & 0.10 & 0.32 \\
\bottomrule
\end{tabular}
\end{table}	

We can see in figure \ref{fig:RD}.(c) the relative difference of the scaled frequency for the Kappa case and Maxwellian case $y^\kappa/y^{MB}$ for the electron cyclotron branch where the supraindex $\kappa$ indicates the frequency dependence in the kappa parameter and the supraindex $\text{MB}$ denote the Maxwellian case (when $\kappa \rightarrow \infty$). We are going to maintain this index notation when analyzing the ponderomotive terms for non-thermal plasmas (see below). In that figure, we can notice that the dispersion relation varies appreciably concerning the kappa parameter at the order of $10^{-2}$ for the electron cyclotron waves for $\omega_{pe}/|\Omega_{e}| = 30$ and $\beta_e \sim 10^{-1}$ which are typical values that can be found in space plasmas. For the Upper branch, we have that for overdense plasmas the effect of the kappa distribution is not significant for $\beta_e < 10$ with a relative difference at most of the order of $10^{-4}$. The effect of the kappa distribution for the Upper branch becomes significant for values of the electron plasma beta of the order of $10^2$ for what we have a relative difference of the wave frequency of $10^{-2}$. These high electron plasma beta values can be found on the inner heliosheath or the ring current where the effect of the kappa could be more relevant for this solution branch (see below). Nevertheless, because of its low non-thermal effect in this work, we would focus mainly on the electron cyclotron waves instead of the Upper branch solution. 

We can also deduce that there is a greater influence of the kappa parameter for the electron cyclotron waves (inherent in the presence of the magnetic field) than the Upper Branch solution, hence the non-thermal effect behaves very differently in the presence of a magnetic field. Also, for parallel wave propagation with the background magnetic field, the non-thermal effect gets enhanced with the plasma beta as we can see in figure \ref{fig:RD}.(c). That can be explained because when the thermal pressure is larger than the pressure of the magnetic field the electrons have more freedom to yield to thermal effects and escape the magnetic field confinement.  \\

\subsection{Ion cyclotron waves}

In the case that we are considering wave frequencies comparable to the ion plasma frequency we will have to consider the effect of the ion in the dispersion relation. Also, if we neglect terms of the order of $m_e/m_i < 10^{-3}$, we consider a quasi-neutral plasma and the left-handed mode of propagation then we have that
\begin{equation}
    \varepsilon_0 = 1 -  \left(\frac{c}{c_A}\right)^2 \frac{1}{(y - 1)}, \qquad \delta = \frac{\beta_i}{2}\left(\frac{\kappa}{\kappa-3/2}\right)\frac{y}{(y - 1)^3}.
    \label{eq:varepsilonColdAW}
\end{equation}
For this case, the scaled frequency is given by $y = \omega/\Omega_i$. Also, $c_A = B_0/\sqrt{4\pi \rho}$ is the Alfvén speed with $\rho = m_e n_e + m_i n_i \approx m_i n_i$ the plasma density and $\beta_e = 8\pi n_i k_B T_i/B_0^2$ is the ion plasma beta. As in the electron cyclotron waves, to accomplish our approximation we are going to work with the range of values for $y$ given by table \ref{tab:RangeOfValidity}. Since, in space plasmas we have typical values of the Alfvén speed of $c_A/c \sim 10^{-3}$ or $\sim 10^{-4}$  \citep{Bourouaine_Alexandrova_Marsch_Maksimovic_2012,Kim_Kim_Kwon_2018} then $|\varepsilon_0| \gg 1$ and we can use the following approximation of the dielectric component for ion cyclotron waves $\varepsilon_0 \approx -\left(c/c_A\right)^2/(y - 1)$. \\

\section{Ponderomotive Force}

The ponderomotive force is a non-linear phenomenon induced by the interaction of a high-frequency field with the plasma in a slow time-scale motion concerning the carrier frequency of the wave \citep{kentwell_time-dependent_1987}. The purpose of this work is to study how the non-thermal effect described by the Kappa distribution impacts the non-linear slow time-scale interaction of magnetized plasmas with electromagnetic fields. Currently, there is a great diversity of formalism from where the ponderomotive force has been derived which has been extensively studied \citep[see][]{kentwell_time-dependent_1987}. In this discussion, we are going to use the Washimi and Karpman ponderomotive force and the previous results for the dispersion relations for low-temperature plasmas characterized by Kappa distributions. This formalism can be obtained either from a stress tensor or fluid formalism for a temporally dispersive and non-absorbing medium \citep{washimi_ponderomotive_1976,karpman_ponderomotive_1982}. Nevertheless, due to the fluid character of its derivation, this force does not work near resonances, where we know that for magnetized plasmas the ponderomotive force produced by cyclotron waves can inject a large amount of energy in the plasma \citep[see][]{lundin_ponderomotive_2007}. In this regime the fluid expression of the ponderomotive force diverges, hence the damping of the wave must be considered to counteract this effect. For this approach, a kinetic formalism must be used, which we will leave for future research.

In the presence of a background magnetic field the ponderomotive force of Washimi and Karpman $\mathbf{f}_{\text{WK}}$, due to the electromagnetic field  $\bar{\mathbf{E}}(\mathbf{r},t)\{=(1/2)[\mathbf{E}(\mathbf{r},t)e^{-i\omega t} + \mathbf{E}^*(\mathbf{r},t)e^{-i\omega t}]\}$, would depend in a term $\mathbf{f}_{(s)}$ associated to the spatial variation of the electric field magnitude, a term $\mathbf{f}_{(t)}$ associated to the temporal variation of the electric field magnitude, a term $\mathbf{f}_{(m)}$ associated with the magnetically induced moment current, and a term $\mathbf{f}_{(MMP)}$ associated with the spatial variation of the background magnetic field:
\begin{equation}
    \mathbf{f}_{\text{WK}} = \mathbf{f}_{(s)} + \mathbf{f}_{(t)} + \mathbf{f}_{(m)} + \mathbf{f}_{MMP}.
    \label{eq:FPwashimikarpman}
\end{equation}
For this particular case, we can express the spatial factor of the ponderomotive force as follows \citep{washimi_ponderomotive_1976}

\begin{equation}
    \mathbf{f}_{(s)} = \frac{1}{16\pi}(\varepsilon_\pm - 1)\nabla |\mathbf{E}|^2.
    \label{eq:FPespacial}
\end{equation}
 If we also suppose that the magnitude of the electric field varies slowly in our time and space scales it can be deduced that the temporal-variation part of the ponderomotive force becomes \citep{washimi_ponderomotive_1976}
\begin{equation}
    \mathbf{f}_{(t)} = \frac{\mathbf{k}}{16\pi\omega^2} \frac{\partial \omega^2 (\varepsilon_\pm - 1)}{\partial \omega}\frac{\partial |\mathbf{E}|^2}{\partial t}.
    \label{eq:FPtemporal}
\end{equation}
Notice that to compute the partial derivative of $\varepsilon_\pm$ in the temporal term of the ponderomotive force we have to use the dielectric tensor given by $\varepsilon_{\pm}(\omega,k) = \varepsilon_{0\pm}(\omega) - \frac{k^2c^2}{\omega^2}\delta_{\pm}(\omega)$ before we use the dispersion relation given by $\varepsilon = k^2c^2/\omega^2$. \\

For this case, we can express the ponderomotive force associated with the currents induced by the ponderomotive magnetic moment as \citep{karpman_ponderomotive_1982}:

\begin{equation}
        \mathbf{f}_{(m)} = \mathbf{B}_0 \times (\nabla \times \mathbf{M}) = B_0 \nabla_\perp M,
        \label{eq:FPm}
\end{equation}
where $M = (1/16\pi)(\partial \varepsilon_\pm/\partial B_0)|\mathbf{E}|^2$ is the nonlinear magnetic moment produced by the ponderomotive force, $\mathbf{B}_0 = B_0 \mathbf{\hat{z}}$ is the background field and $\nabla_\perp = \nabla - \mathbf{\hat{z}}\frac{\partial }{\partial z}$ is the gradient transverse to the magnetic field. \\

Finally, we have the term associated with the spatial variation of the background magnetic field which is called {\it magnetic moment pumping}, and which can be considered as the force applied to a magnetic dipole due to an external magnetic field \citep{lundin_ponderomotive_2007}
\begin{equation}
        \mathbf{f}_{(MMP)} = M\nabla B_0.
        \label{eq:FPmmp}
\end{equation}

Now let us analyze the factors $f_{(s)}\left\{=(1/8\pi)(\varepsilon_+-1)\right\}$ and $f_{(t)}\{= (k/16\pi \omega^2)[d \omega^2(\varepsilon_+ - 1)/d \omega] \}$ that accompany the spatial and temporal variations of the magnitude of the electric field in the ponderomotive force, and the magnetic moment $M$ due to the propagation of electron cyclotron waves and ion cyclotron waves described by Kappa distributions. \\

\subsection{Spatial ponderomotive force factor} 

For electron cyclotron waves we can deduce using our results for the dielectric eigenvalue $\varepsilon_+$ that the factor $f^\kappa_{(s)}$ that accompanies the spatial variation in the ponderomotive force for plasmas characterized by Kappa distributions is given by (see appendix B for more details|)
\begin{equation}
    f_{(s)}^\kappa = \frac{1}{16\pi} \left(\frac{\omega_{pe}}{|\Omega_e|}\right)^2\frac{1}{y(1-y)}+\frac{\beta_{e}}{32\pi}\left(\frac{\kappa}{\kappa-3/2}\right)\left(\frac{\omega_{pe}}{|\Omega_e|}\right)^2 \frac{1}{(1-y)^4},
    \label{eq:FPespacial_nonthermal}
\end{equation}
with $y = \omega/|\Omega_e|$. Here we can notice that if we neglect our finite temperature correction; i.e. we put $\beta_e = 0$ in equation (\ref{eq:FPespacial}), then we recover the spatial term of the ponderomotive force for electron cyclotron waves in cold plasmas \citep{lundin_ponderomotive_2007}. Also, if we do $|\Omega_e| = 0$, considering that $\beta_e = (\omega_{pe}^2/|\Omega_e|^2)(\alpha_e^2/c^2)$ we recover the expression deduced for the spatial ponderomotive term in the unmagnetized case \citep{Espinoza-Troni_A_Asenjo_Moya_2023} (except for that we have considered here that $|\varepsilon_0| \gg 1$).\\

We can notice that the spatial term of the ponderomotive force is negative for the Upper branch solution ($\omega > |\Omega_e|$) and positive for the electron cyclotron waves ($\omega < |\Omega_e|$). Therefore, the spatial term of the ponderomotive force will push the plasma against (in favor) of the gradient of the amplitude of the wave for the Upper branch (electron cyclotron) waves. Nevertheless, for overdense plasmas, as is typical in near-Earth spatial environments (see section 4) the frequency for the Upper branch solution is very large compared to the frequency of the electron cyclotron waves for the same wavenumber. Hence, the magnitude of the spatial term of the ponderomotive force for the Upper branch waves is negligible compared to the electron cyclotron wave term. Indeed, for $\omega_{pe}/|\Omega_e| = 30$, $x = 0.3$ and $\beta_e = 0.9$ we have that the ratio of the frequency for the Upper branch and electron cyclotron waves is of the order of $10^{-5}$. Also, for the Upper branch, the ponderomotive force decreases with the wavenumber contrary to what happens for electron cyclotron waves that approach the resonance. \\ 


We can notice that we have a resonance for $\omega = 0$, where our results are not valid for electron waves since we would have to consider the effect of the ions. Also our equation (\ref{eq:FPespacial_nonthermal}) shows a resonance in $\omega = |\Omega_e|$ nevertheless this is not valid since we have considered as an approximation that $\delta \ll 1$. Hence, when we approach the resonances we have to use the dispersion relation given by (\ref{eq:varepsilon}). The same observation can be extended to the other ponderomotive force terms. To study that characteristic we would also have to extend the formalism to include the damping of the wave. Also, the spatial term of the ponderomotive force has a minimum which is enhanced with the non-thermal effect and is shifted to lower frequencies. \\

\begin{figure*}[ht]
\centering
{\includegraphics[width=0.75\linewidth]{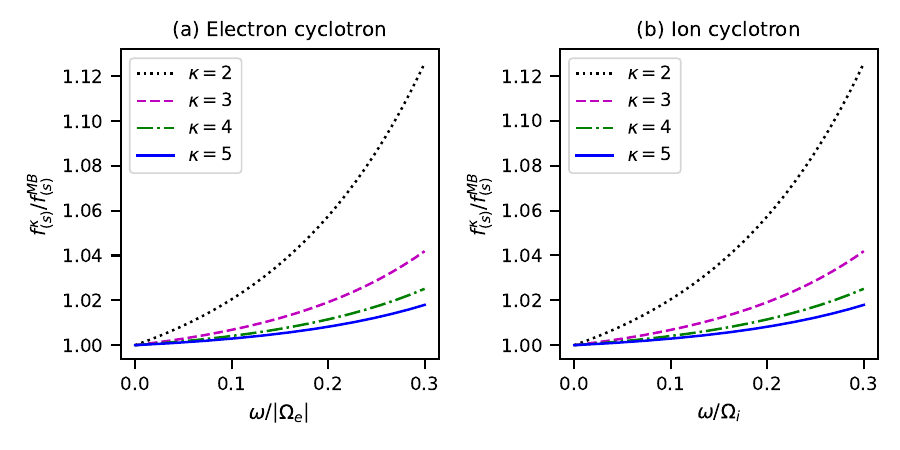}}
\caption{Ratio of the Kappa and Maxwellian ponderomotive spatial factor $f^\kappa_{(s)}/f^{\text{MB}}_{(s)}$ (a) for Electron cyclotron waves as function of $\omega/|\Omega_e|$ with $\beta_e = 0.1$. (b) for ion cyclotron waves as function of $\omega/\Omega_i$ with $\beta_i = 0.1$.}
\label{fig:FP_spatial}
\end{figure*}

Figure \ref{fig:FP_spatial}.(a) shows the ratio of the Kappa and Maxwellian ponderomotive spatial factor $f_{(s)}^\kappa/f_{(s)}^\text{MB}$ as a function of the scaled wavenumber $\omega/|\Omega_e|$ for the electron cyclotron waves. As it is seen, the magnitude of the ponderomotive force for the Kappa distributions is greater compared to the Maxwellian distribution case and gets enhanced with the decrease of the kappa parameter. Besides, when the scaled wavenumber decreases, both magnitudes are equal so the non-thermal effect is canceled. It is relevant to note that the relative difference of the Kappa and Maxwellian spatial factor is at the order of $10^{-2}$ for the electron cyclotron waves for $\beta_e = 0.1$ and its dependence in $\omega_{pe}/|\Omega_e|$ is very low (indeed, for our overdense plasmas approximation $|\varepsilon_0| \gg 1$ is independent of $\omega_{pe}/|\Omega_e|$), unlike the unmagnetized case where we had for non-relativistic velocities of $\alpha_e/c \sim 10^{-2}$ no more than a relative difference of the order of $10^{-5}$ \citep{Espinoza-Troni_A_Asenjo_Moya_2023}, so we can deduce that the effect of the kappa parameter in the ponderomotive force becomes more relevant when we include the effect of a background magnetic field in the plasma. Also, the non-thermal effect is larger for overdense plasmas. As we will see below these parameters of the plasma beta and the frequencies ratio can be found in the solar wind, the near magnetotail, and the plasma sheet for we can conclude that in these environments the effect of the kappa parameter in the electron cyclotron ponderomotive force is significant even for low-frequency waves. Also, we expect that the effect of the kappa parameter can be more notorious for larger values of the frequency and the plasma beta, for which we will have to relax our approximation.\\

For ion cyclotron waves we can deduce using our result for the dielectric eigenvalue $\varepsilon_+$ that the factor $f^\kappa_{(s)}$ that accompanies the spatial variation in the ponderomotive force for plasmas characterized by Kappa distributions is given by
\begin{equation}
    f_{(s)}^\kappa = \frac{1}{16\pi}\left(\frac{c}{c_A}\right)^2\frac{1}{(1-y)} + \frac{1}{32\pi}\beta_{i}\left(\frac{\kappa}{\kappa-3/2}\right) \left(\frac{c}{c_A}\right)^2 \frac{y}{(1-y)^4},
\end{equation}
with $y = \omega/\Omega_i$. We can notice that as is the case for electron cyclotron waves, the spatial term of the ponderomotive force for ion cyclotron waves will push the plasma in favor of the gradient of the wave amplitude. For very low frequencies $\omega \ll \Omega_i$ the non-thermal effects are canceled and the spatial ponderomotive factor tends to be constant with an asymptotic value of $(1/16\pi)(c/c_A)^2$. Figure \ref{fig:FP_spatial}.(b) shows the ratio of the Kappa and Maxwellian ponderomotive spatial factor $f_{(s)}^\kappa/f_{(s)}^\text{MB}$ as a function of the scaled wavenumber $\omega/\Omega_i$ for the ion cyclotron waves. We have that the spatial term of the ponderomotive force for the ion cyclotron wave is enhanced with the kappa parameters at the same ratio as the electron cyclotron waves. This was expected because the $\delta$ term has the same form for both modes of propagation. \\

Also, due to the factor $\beta_e$ multiplying $\kappa/(\kappa-3/2)$  the non-thermal effect increases with the plasma beta in the low plasma beta domain that we are analyzing for either electron cyclotron waves and ion cyclotron waves, as it could be expected because the larger the thermal effect to the magnetic field the less confined are the electrons in the path described by the magnetic field as we discussed above when we analyzed the dispersion relation. Therefore, it is expected that the effect of the kappa parameter can be enhanced for larger values of the plasma beta as what can be found in the magnetic holes produced by the ponderomotive force of the steepened Alfvén waves in the solar wind, for what we can have values of $\beta \sim 1$ \citep{dasgupta_2003}. Also, we have that in the dayside magnetosheath we have values for the $\beta_i$ of ions between 1 and 13 with an average of $\beta_i \sim 3.5$, for what we would have to use the full expression of the generalized plasma dispersion function for ions and we could use our approximation for electrons whose temperature is one order of magnitude lower. Nevertheless when the magnetic shear between the magnetopause and the magnetosheath is low, this last presents a transition layer to the magnetopause where the plasma beta can be lower than 1 for both species reaching values of $\beta_i \sim 0.4$ for where our approximation could be used including ions \citep{Phan_Pashmann}. Also, we can notice that for $\beta = 0.1$, a scaled frequency of $0.3$ and $\kappa = 2$ the spatial ponderomotive factor for both electron cyclotron waves and ion cyclotron waves is $12\%$ larger than for Maxwellian plasmas. We have then a significant impact of the kappa parameter in the magnitude of the spatial term of the ponderomotive force, and therefore its implications in space phenomena could be useful as a tool to make measurements of the kappa parameter and to get a better understanding of the velocity distribution in space environments.

\subsection{Temporal ponderomotive force factor}

Here we analyze the influence of the non-thermal effect in the factor $f_{(t)}^\kappa$ that accompanies the temporal variation of the wave amplitude in the ponderomotive force for electron cyclotron waves and ion cyclotron waves. We have for the electron cyclotron waves that
\begin{equation}
    f_{(t)}^\kappa = \frac{1}{16\pi c}\left(\frac{\omega_{pe}}{|\Omega_e|}\right)^3\frac{1}{y^{3/2}(1-y)^{5/2}}\left[1+ \frac{\beta_{e}}{4}\left(\frac{\kappa}{\kappa-3/2}\right)\frac{(3+4y)y}{(1-y)^3}  \right],
    \label{eq:PFtemporalEC}
\end{equation}
with $y = \omega/|\Omega_e|$. From the previous equation, it follows that the temporal factor of the ponderomotive force is non-zero, even if we consider zero temperature (i.e for $\beta_e=0$), unlike what happens for unmagnetized plasmas \citep{Espinoza-Troni_A_Asenjo_Moya_2023} whose temporal ponderomotive term is canceled for cold plasmas. \\

\begin{figure*}[ht]
\centering
{\includegraphics[width=0.75\linewidth]{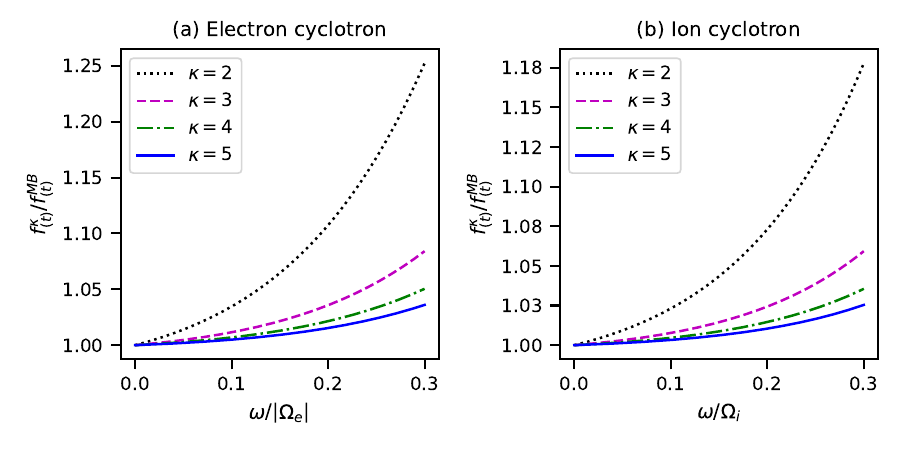}}
\caption{ Ratio of the Kappa and Maxwellian ponderomotive temporal factor $f^\kappa_{(t)}/f^{\text{MB}}_{(t)}$ (a) for Electron cyclotron waves as function of $\omega/|\Omega_e|$ with $\beta_e = 0.1$. (b) for ion cyclotron waves as function of $\omega/\Omega_i$ with $\beta_i = 0.1$.}
\label{fig:FP_temporal}
\end{figure*}

We can notice that the temporal factor of the ponderomotive force is positive for the electron cyclotron waves. The direction of the temporal term of the ponderomotive force depends on the direction of the propagation of the wave and the temporal variation of the wave amplitude as can be seen more clearly in equation (\ref{eq:FPtemporal}). This force will push the plasma in favor (against) the wave propagation direction for the temporal increase (decrease) of the wave amplitude. As is the case of the ponderomotive spatial term, for overdense plasmas the magnitude of the temporal term of the ponderomotive force for the Upper branch waves is negligible compared to the electron cyclotron wave term. Indeed, if we use direct expression (\ref{eq:FPtemporal}) (since our approximation $\varepsilon_0 \gg 1$ is not valid for the Upper Branch solution) for $\omega_{pe}/|\Omega_e| = 30$, $x = 0.3$ and $\beta_e = 0.9$ we have that the ratio of the frequency for the Upper branch and electron cyclotron waves is of the order of $10^{-9}$. Also, for the Upper branch, the ponderomotive force decreases with the wavenumber contrary to what happens for electron cyclotron waves that approach the resonance.   \\


Also in this case we have a minimum that decreases with non-thermal effects and is shifted to lower frequencies.  Figure \ref{fig:FP_temporal}.(a) shows the ratio of the Kappa and Maxwellian ponderomotive temporal factor for right-handed electron cyclotron waves $f_{(t)}^\kappa/f_{(t)}^\text{MB}$ as function of the scaled frequency $\omega/|\Omega_e|$ for $\beta_e = 0.1$. For overdense plasmas, we can notice in this figure that the ponderomotive temporal factor for electron cyclotron waves is larger for Kappa-distributed plasmas than for Maxwellian plasmas in an order of magnitude of $\sim 10^{-1}$. We can notice also that the ponderomotive temporal factor is proportional to $(\omega_{pe}/|\Omega_e|)^3$.  \\

Now we are going to analyze the ponderomotive temporal factor for ion cyclotron waves which is given by
\begin{equation}
    f_{(t)}^\kappa = \frac{1}{16\pi c}\left(\frac{c}{c_A}\right)^3 \frac{(2-y)}{(1-y)^{5/2}}\left[1+ \frac{\beta_{i}}{4}\left(\frac{\kappa}{\kappa-3/2}\right)\frac{(4+3y)y}{(2-y)(1-y)^3} \right],
\end{equation}
with $y = \omega/\Omega_i$. Also, we can notice that for very low frequencies $\omega \ll \Omega_i$ the temporal ponderomotive factor tends to a constant asymptotic value given by $(1/8\pi c)(c/c_A)^3$, where the finite temperature effects are canceled. Figure \ref{fig:FP_temporal}.(b) shows the ratio of the Kappa and Maxwellian ponderomotive temporal factor for ion cyclotron waves $f_{(t)}^\kappa/f_{(t)}^\text{MB}$ as a function of the scaled frequency $\omega/\Omega_i$ for $\beta_i = 0.1$. We can notice in this figure that the ponderomotive temporal factor for ion cyclotron waves is larger for Kappa-distributed plasmas than for Maxwellian plasmas in an order of magnitude of $\sim 10^{-1}$. Also, for both electron cyclotron and ion cyclotron waves the non-thermal effect increases with the plasma beta as we expected due to a less magnetic confined plasma as we discussed above.  \\

Finally, we can notice that for $\beta = 0.1$, a scaled frequency of $0.3$ and $\kappa = 2$ the temporal ponderomotive factor is $25\%$ and $18\%$ larger than for Maxwellian plasmas for electron cyclotron waves and ion cyclotron waves respectively. Therefore, the non-thermal effect can be significant when evaluating the temporal term of the ponderomotive force even for low-temperature plasmas. On the other hand, this term can be relevant when we approach the resonances for what we would have to consider the damping of the wave, and therefore its amplitude would have a temporal variation. For that case terms of the order of $\delta^2$ could not be neglected, unless we consider low damping far from resonances. As we would see below, this temporal ponderomotive term also acts to perturb the slow time scale background magnetic field $B_0$.   \\

\subsection{Magnetic moment of the ponderomotive force}

In this section, we analyze the ponderomotive magnetic moment factor $M = (1/16\pi)(\partial \varepsilon / \partial B)|E|^2$ produced by the wave propagation in Kappa distributed plasmas which is responsible for the induced current force and the MMP force of equations (\ref{eq:FPm}) and (\ref{eq:FPmmp}). We obtain that the magnetic moment for certain kappa value $M^\kappa$ for the electron cyclotron waves is given by
\begin{equation}
M^\kappa = -\frac{|E|^2}{16\pi B_0}\left(\frac{\omega_{pe}}{|\Omega_e|}\right)^2\frac{1}{y(1-y)^2}\left[1 +  \frac{3}{2}\beta_{e}\left(\frac{\kappa}{\kappa-3/2}\right)\frac{y}{(1-y)^3} \right],
\end{equation}
where $y = \omega/|\Omega_e|$. As is the case for the previous ponderomotive force terms, for overdense plasmas the magnitude of ponderomotive magnetic moment for the Upper branch waves (without considering $\varepsilon_0 \gg 1$) is negligible compared to the electron cyclotron wave term. On the other hand, for ion cyclotron waves we have that
\begin{equation}
M^\kappa = -\frac{|E|^2}{8\pi B_0}\left(\frac{c}{c_A}\right)^2 \frac{(2-y)}{(1-y)^2}\left[1 + \frac{3}{2}\beta_{i}\left(\frac{\kappa}{\kappa-3/2}\right)\frac{y}{(2-y)(1-y)^3}\right],
\end{equation}
where $y = \omega/\Omega_i$. We can notice that when $\omega \ll \Omega_i$ the ponderomotive magnetic moment tends to an asymptotic value given by -$(|E|^2/8\pi)(c/c_A)^2$ and the finite temperature effects are canceled. We can notice that the ponderomotive magnetic moment is negative for both the electron cyclotron waves and the ion cyclotron waves. Therefore, the MMP term of the ponderomotive force will push the plasma against the gradient of the amplitude of the background magnetic field. For this reason, this force is responsible for the acceleration of ions in the polar cusps, where the ponderomotive force pushes the plasma out of the polar regions \citep{Li_Temerin_1993,guglielmi_ponderomotive_2001}. Also, we can notice that in both the cold and low-temperature plasma case as we get closer to the resonance the ponderomotive magnetic moment tends to be larger in the negative direction. \\

\begin{figure*}
\centering
{\includegraphics[width=0.8\linewidth]{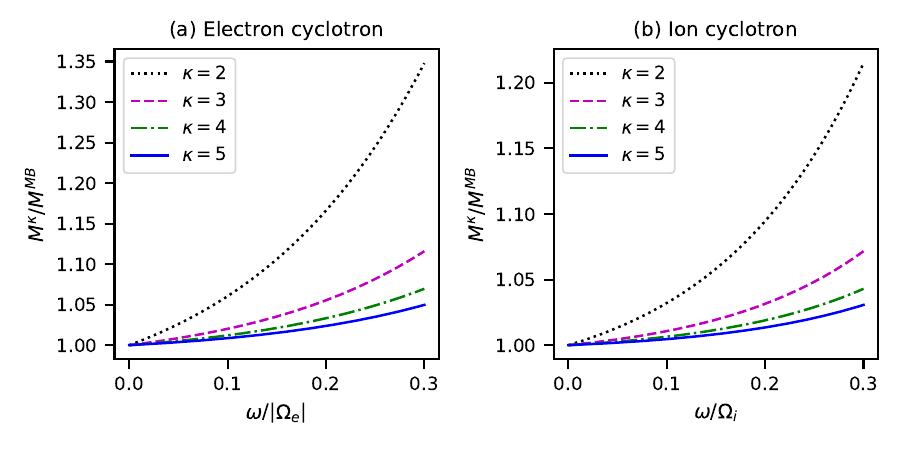}}
\caption{Ratio of the Kappa and Maxwellian ponderomotive magnetic moment $M^\kappa/M^{\text{Mb}}$ (a) for Electron cyclotron waves as function of $\omega/|\Omega_e|$ with $\beta_e = 0.1$. (b) for ion cyclotron waves as function of $\omega/\Omega_i$ with $\beta_i = 0.1$.}
\label{fig:PF_mm}
\end{figure*}

Figures \ref{fig:PF_mm}.(a) and \ref{fig:PF_mm}.(b)  shows the relative ratio of the ponderomotive magnetic moment $M$ for Kappa and Maxwellian distributed plasmas as a function of the scaled frequency for electron cyclotron waves and ion cyclotron waves. We can notice that for both modes of propagation in the range of frequencies under consideration, the ponderomotive magnetic magnitude is larger for the Kappa case than the Maxwellian case. Also, we can see that even for low values of the plasma beta the non-thermal effect is very significant for both the electron cyclotron waves and ion cyclotron waves. Indeed for $\beta = 0.1$, we have that the ponderomotive magnetic moment for the Kappa distributed plasma can be $35\%$ and $20\%$ larger than for the Maxwellian distributed plasma for $\kappa = 2$ for electron and ion cyclotron waves respectively. The same behavior can be seen in the spatial and temporal terms of the ponderomotive force where for $\beta = 0.1$, $\kappa = 2$, and a scaled frequency of $0.3$ we have for the electron cyclotron waves and ion cyclotron waves that at least $f_{(s)}^\kappa/f_{(s)}^\text{MB} \approx 1.12$ and $f_{(t)}^\kappa/f_{(t)}^\text{MB} \approx 1.18$. \\

This result is very important because it tells us that even for nearly cold magnetized plasmas and for low frequencies the effect of the kappa parameter in the ponderomotive force can not be neglected, therefore it must be considered when it is applied to study wave-plasma interactions in space phenomena occurring in a large diversity of low collision space plasma environments where the Kappa distribution is usually present as we will see below.

\subsection{Nonlinear magnetic field perturbation due to the ponderomotive force}

According to the work of Washimi and Watanabe \citep{washimi_magnetic_1977} a slowly varying magnetic field $\mathbf{B}_2$ is generated by the ponderomotive force of the electromagnetic wave that is slowly varying in time. From the balance of the ponderomotive force with the slowly varying electromagnetic field in the electron-fluid equation of motion of an unmagnetized and homogeneous plasma, it follows that the induced magnetic field is given by
\begin{equation}
    \mathbf{B}_2(\mathbf{r},t) = -\frac{c}{16\pi n_e e \omega^2}\frac{\partial [\omega^2(\varepsilon - 1)]}{\partial \omega}\nabla\times (\mathbf{k}|\mathbf{E}|^2),
    \label{eq:WKponderomotiveMagneticField}
\end{equation}
where $e$ is the electron charge. This induced magnetization has been widely studied as a mechanism of self-generated magnetic field. It has been analyzed for different plasma conditions as can be for example relativistic electron plasmas \citep{Qi_Zhan_Liu_Yang_2023} or dense plasmas \citep{shukla_generation_2010} due to its importance among others in phenomena associated with pulsar magnetospheres or compact astrophysical objects respectively. Kim and Jung have calculated the Washimi and Karpman ponderomotive magnetic field for the non-thermal electrostatic case \citep{kim_nonthermal_2009} and it also was analyzed for the electromagnetic case in \citep{Espinoza-Troni_A_Asenjo_Moya_2023} where it was obtained that the non-thermal effect of the Kappa distributions enhances the induced magnetization due to the electromagnetic ponderomotive interactions in unmagnetized plasmas. In this subsection, we are going to analyze the induced Washimi and Karpman ponderomotive magnetization for non-thermal magnetized plasmas. This would lead us to advance in the understanding of the effect of the ponderomotive force in the non-linear perturbation of the background magnetic field in space environments. In our work, we are considering plasmas with electrons and ions, but we can still use the Washimi and Karpman induced magnetic field considering the ponderomotive force only for electrons, since $m_e/m_i \ll 1$.  \\
\begin{figure}[h!]
\resizebox{0.91\hsize}{!}{\includegraphics{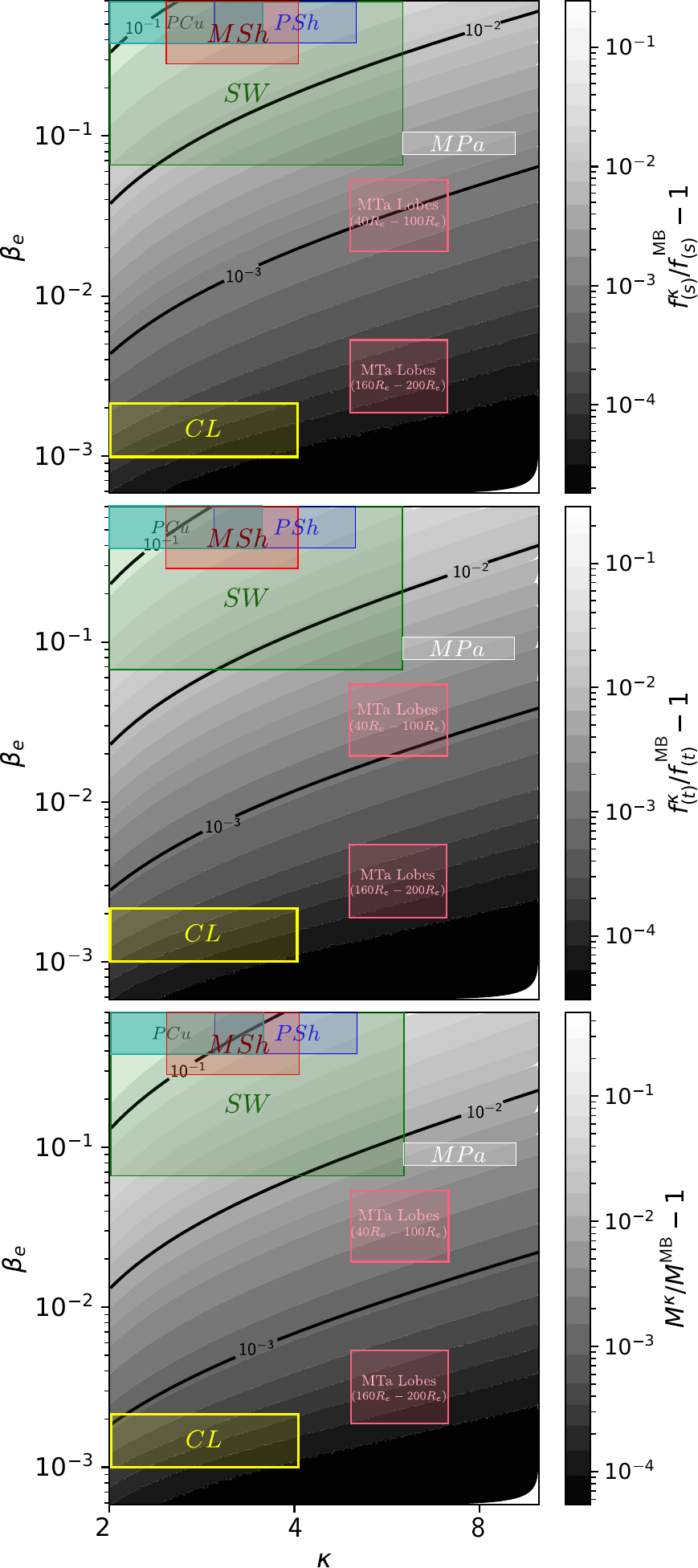}}
\caption{Relative difference of the different terms of the ponderomotive force between the Kappa and Maxwellian case, as a function of $\beta_e$ and $\kappa$. The ordinate axis and the color bar are in the logarithmic scale in base 10, and the abscissa axis is in the logarithmic scale in base 2. The abbreviations are explained in table \ref{tab:PF_ParametersTable}.}
\label{fig:HeatMapsPF.pdf}
\end{figure}

\renewcommand\arraystretch{1.4}
\renewcommand{\tabcolsep}{6pt}
\definecolor{LightCyan}{rgb}{0.88,1,1}
\definecolor{Gray}{gray}{0.9}
\definecolor{LightGray}{gray}{0.93}

\newcolumntype{g}{>{\columncolor{LightGray}}c}
\newcolumntype{C}{>{\columncolor{LightCyan}}l}

\begin{table*}
\caption{Evaluation of the different terms of the ponderomotive force and the induced Washimi and Karpman magnetic field $B_2$. The relative difference of the ponderomotive force terms for the Kappa and Maxwellian case is expressed in percentage for a range of typical values of the parameters $\omega_{pe}/|\Omega_{e}|$, $\beta_e$ and $\kappa$, with $\omega/|\Omega_e| = 0.1$ for a variety of spatial condition showed by their abbreviations explained below. Coronal loops(CL); Solar wind (SW); Magnetosheath (MSh); Magnetopause (Mpa); Magnetotail lobes (MTaLb); Plasmasheet (PSh); Polar cusps (PCu); Inner heliosheath (IH). The numbers in parentheses represent the magnitude order. The ponderomotive temporal factor and the induced Washimi and Karpman magnetization share the same column. }
\label{tab:PF_ParametersTable}
\begin{small}
\begin{tabular}{C|ggggggC}
\toprule
\rowcolor{white}
Environments & $\omega_{pe}/|\Omega_{e}|$ & $\beta_e$ & $\kappa$ & $f_{(s)}^\kappa/f_{(s)}^\text{MB}-1$ & $f_{(t)}^\kappa/f_{(t)}^\text{MB}-1$ & $M^\kappa/M^\text{MB}-1$ & References \\ 
\rowcolor{white}
 &  & & &  & $B_2^\kappa/B_2^\text{MB} - 1$  &  \\ 
\hline
CL  & $1.0-1.6$ & $4-20\,\,(-4)$ & $1.8-4.0$ & $0.002\%-0.072\%$ & $0.003\%-0.117\%$ & $0.005\%-0.205\%$ & 1,2 \\
\rowcolor{white}
SW (0.25 AU) & $40-90$ & $0.07-0.7$ & $2-6$ & $0.2\%-18\%$ & $0.3\%-23\%$ & $0.5\%-38\%$ & 3,4,5,6 \\
SW (1 AU) & $100-140$ & $0.6-2.0$ & $2-6$ & $1\%-18\%$  & $2\%-23\%$ & $4\%-38\%$ & 3,4,5,6 \\
\rowcolor{white}
MSh & $20-60$ & $0.3-0.7$ & $2.5-4.0$ & $1\%-8\%$ & $2\%-11\%$ & $3\%-19\%$ & 7,8 \\
MPa & $9-12$ & $0.08-0.10$ & $6-9$ & $0.1\%-0.2\%$ & $0.2\%-0.4\%$ & $0.3\%-0.7\%$ & 7,9 \\
\rowcolor{white}
MTaLb & $2-7$ & $ 2-5\,\,(-3)$ & $5-7$ & $0.004\%-0.015\%$ & $0.006\%-0.025\%$ & $0.011\%-0.044\%$ & 10,11 \\ 
\rowcolor{white}
($40R_e - 100R_e$) & & & & & & &   \\  
MTaLb & $10-17$ & $0.02-0.05$ & $5-7$  & $0.04\%-0.15\%$ & $0.06\%-0.25\%$ & $0.11\%-0.44\%$ & 10,11 \\ 
($160R_e-200R_e $) & & & & & & &  \\      
\rowcolor{white}
PSh & $10-40$  & $0.4-1.0$ & $3-5$ & $1\%-5\%$ & $2\%-8\%$ & $3\%-13\%$ & 11,12   \\
\rowcolor{white}
($60R_e - 100 R_e$) & & & & & & &  \\ 
PSh & $30-60$ & $0.6-1.5$ & $3-5$ & $2\%-5\%$ & $3\%-8\%$ & $5\%-13\%$  & 11,12 \\
($160R_e - 200 R_e$) & & & & & & &  \\
\rowcolor{white}
PCu  & $10-50$ & $0.4-10$ & $2.0-3.5$ & $2\%-18\%$ & $3\%-23\%$ & $6\%-38\%$ & 13 \\
IH & $200-1000$ & $30-350$ & $1.6-1.9$ & -- & -- & -- &  14,15 \\
\rowcolor{white}
Ring Current & $9-70$ & $2-200$ & $5.5-6.5$ & -- & -- & -- & 16 \\
Near MTa & $14-48$ & $1.7-5.4$ & $2.4-4.5$ & -- & -- & -- & 17 \\
($7R_e- 20R_e$) & & & & & & & \\
\bottomrule
\end{tabular}
\tablebib{(1)~\citet{Brooks_Warren_Landi_2021}; (2) \citet{Vocks_Mann_Rausche_2008}; (3) \citet{Matteini_Hellinger_Landi_Trávníček_Velli_2011}; (4) \citet{WilsonIII_Stevens_Kasper_Klein_Maruca_Bale_Bowen_Pulupa_Salem_2018}; (5) \citet{Livadiotis_2015}; (6) \citet{Bale_Goetz_Harvey_Turin_Bonnell_Dudoknbsp;denbsp;Wit_Ergun_MacDowall_Pulupa_Andre_etal._2016}; (7) \citet{Phan_Pashmann}; (8) \citet{Ogasawara_Angelopoulos_Dayeh_Fuselier_Livadiotis_McComas_McFadden_2013}; (9) \citet{Kirpichev_Antonova}; (10) \citet{Lui_Krimigis_1983}; (11) \citet{Slavin_Smith_Sibeck_Baker_Zwickl_Akasofu_1985}; (12) \citet{Kletzing_2003}; (13) \citet{Ren_Zong_Fu_Yang_Hu_Zhang_Zhou_Yue_Kistler_Daly_etal._2023}; (14) \citet{Livadiotis_McComas_Funsten_Schwadron_Szalay_Zirnstein_2022}; (15) \citet{Burlaga_Ness_Acuna_2006}; (16) \citet{Pisarenko_Budnik_Ermolaev_Kirpichev_Lutsenko_Morozova_Antonova_2002}; (17) \citet{Runov_Angelopoulos_Gabrielse_Liu_Turner_Zhou_2015}. }
\end{small}
\end{table*}	

Also, we can notice that equation (\ref{eq:WKponderomotiveMagneticField}) is related to the temporal term of the ponderomotive force. Therefore, using equation (\ref{eq:PFtemporalEC}) for the temporal factor of the ponderomotive force for electron cyclotron waves we can deduce the following expression for the magnitude of the induced Washimi and Karpman magnetic field:
\begin{equation}
    B_2^\kappa = \frac{1}{16\pi}\left(\frac{\omega_{pe}}{|\Omega_e|}\right)^3\frac{1}{y^{3/2}(1-y)^{5/2}}\left[1+\frac{\beta_e}{4}\left(\frac{\kappa}{\kappa-3/2}\right)\frac{(3+4y)y}{(1-y)^3}\right]\frac{|E|^2}{n_e e L},
\end{equation}
where $L$ is the scale length of the intensity of the field and $y = \omega/|\Omega_e|$. Using this we can calculate the scaled electron cyclotron frequency $\omega_{ce}/|\Omega_e|$ generated by the induced magnetic field: 
\begin{equation}
    \omega_{ce}/|\Omega_e| = M_p(\kappa,y)\left(\frac{\omega_{pe}}{|\Omega_e|}\right)^4\left(\frac{u_e^2}{cL\omega_{pe}}\right),
\end{equation}
where $u_e = e|E|/m_e\omega_{pe}$ is the electron quiver velocity \citep{kourakis_magnetization_2006} and we have defined the Karpman–Washimi ponderomotive magnetization $M_p(\kappa,y)$ as in \citet{kim_nonthermal_2009}:
\begin{equation}
    M_p(\kappa,y) =  \frac{1}{4}\frac{1}{y^{3/2}(1-y)^{5/2}}\left[1+\frac{\beta_{e}}{4}\left(\frac{\kappa}{\kappa-3/2}\right)\frac{(3+4y)y}{(1-y)^3}\right].
\end{equation}
We can notice that this term has the same form and qualitative behavior as the temporal term of the ponderomotive force for electron cyclotron waves in equation (\ref{eq:PFtemporalEC}) already analyzed. Therefore the induced magnetization would increase with the kappa parameter. Hence, the non-thermal effect of the Kappa distributions enhances the induced magnetization due to the electromagnetic ponderomotive interactions in magnetized plasmas with parallel propagation. Also, it has a minimum for a certain value of the frequency. We also have to notice also, that for our deduction to be valid, we must have $\omega_{ce}/|\Omega_e| \ll 1$. Then, this expression is valid for waves amplitudes with slow spatial variation that satisfy  $(\omega_{pe}/|\Omega_e|)^4\left(u_e^2/cL\omega_{pe}\right) \ll 1$. On the other hand, we would have to extend the Washimi and Karpman formalism for a non-perturbative ponderomotive force deduction.    \\

In the non-relativistic limit we can calculate the total radiated power $P$ average in one period produced by the gyromotion of the charges by the induced Washimi and Karpman magnetic field \citep{na_temperature_2009}, using the Larmor formula \citep{jackson_classical_1975}, is $P = \frac{2}{3}\frac{e^2r_L^2}{c^3}M_p(\kappa,y)^4\left(\frac{\omega_{pe}}  {|\Omega_e|}\right)^{16}\left(\frac{u_e^2}{cL\omega_{pe}}\right)^4$, where $r_L$ is the Larmor radius. Because the induced magnetization increases with the decreasing of the spectral index, it is expected that the non-thermal effect enhances the total energy radiated in magnetized non-thermal plasmas. This result serves as a diagnostic tool for non-thermal magnetized space plasmas. \\

\section{Ponderomotive force in nonthermal plasmas for different space conditions}

To contextualize and specify our results we have calculated the relative difference between Kappa and Maxwellian ponderomotive force factors for different space environments due to electron cyclotron waves. These calculations were made in a range of typical values of the plasma parameters for each space environment and for frequencies low compared with the electron gyrofrecuency with $\omega/|\Omega_e| = 0.1$ for which our approximation ceases to be valid for $\beta > 0.7$. These results are displayed in table \ref{tab:PF_ParametersTable} and shown graphically in figure \ref{fig:HeatMapsPF.pdf}. The values of the plasma beta and the ratio between the plasma frequency and gyrofrequency were calculated using the observations of the plasma electron density, temperature, and the strength of the background magnetic field analyzed in the references mentioned in the table. It is worth noting that these values are a rough estimation, based on the observational articles given in the table. For the environments where the plasma beta can be larger than $0.7$, we evaluated the relative difference of the ponderomotive force terms up to this value, to be coherent with our approximation. Although these results are a general estimation and strongly depend on space weather conditions they serve to give us a concrete idea of the impact of the nonthermal effects in the ponderomotive force and its wave-particle phenomena associated with it in a wide variety of space environments. We can see in figure \ref{fig:HeatMapsPF.pdf} that the effect of the Kappa distribution in the three terms of the ponderomotive force analyzed before can be very significant for the solar wind, the magnetosheath, the plasmasheet, and the polar cusp in the presence of interplanetary (IP) shocks having a relative difference between the Kappa and Maxwellian case of the order of $10^{-2}$ or $10^{-1}$. Indeed in the solar wind between $0.25\text{AU}$ and $0.1\text{AU}$ where we have low values of the kappa parameter between $2$ and $6$ we have that the ponderomotive force can be $18\%$, $23\%$ and $38\%$ larger for the kappa distribution case for the spatial ponderomotive term, the temporal ponderomotive term and the ponderomotive magnetic moment respectively. Therefore, in the near-Earth solar wind at $1\text{AU}$, the MMP term of the ponderomotive force, which will accelerate the plasma in the decreasing direction of the magnetic field strength along the interplanetary magnetic field (IMF) lines could be $38\%$ larger for values of the frequency far from the resonances. Nevertheless, it can have plasma beta values larger than $1$ for what we would have to relax out approximation to evaluate the ponderomotive force in these cases.  In the case of the polar cusp, we have evaluated the ponderomotive force relative differences in the occurrence of an IP shock which are associated with diamagnetic cavities (DMC) due to the increase of plasma density and pressure as a result of the magnetosheath plasma
accumulation in the cusp region, which is usually filled with energetic particles as it is analyzed in \citep{Ren_Zong_Fu_Yang_Hu_Zhang_Zhou_Yue_Kistler_Daly_etal._2023} where it is mentioned that among other theories it has been proposed that the source of the energetic particles are due to wave-particle interaction mechanisms. This environment is also interesting because, as we have said before, in the polar cusps region the MMP ponderomotive force term is responsible for the acceleration and escape of ions along the open magnetic field lines, which is known as polar wind \citep{Li_Temerin_1993,Miller_Rasmussen_Combi_Gombosi_Winske_1995,guglielmi_ponderomotive_2001,Guglielmi_2007}. Our results in table \ref{tab:PF_ParametersTable} show that for this environment where the kappa parameter can have values of $\sim 2$ the absolute value of the relative difference of the ponderomotive force under the beta range of our approximation can be of $18\%$, $23\%$ and $38\%$ for the spatial, temporal and magnetic moment terms. These results tell us that is essential to consider the non-thermal effects of the ponderomotive force when studying these phenomena. 

For other space environments as the coronal loops in the solar corona, the magnetotail lobes (for distances relative to Earth larger than $40 R_e$) and the magnetopause where the magnetic pressure is very low in comparison with the plasma pressure, the nonthermal effect is negligible, for what we have a relative difference of the ponderomotive force terms of the order of $10^{-4}$ or even $10^{-5}$. Hence in these environments under typical conditions, the Maxwelian ponderomotive force terms can be used without issues, unless we approach the resonances. Nevertheless for the near magnetotail (between $7R_e$ and $20R_e$) where the temperature is bigger than for larger distances and therefore the plasma beta is greater than $1$, we would have to extend our results, but our analysis in the sections above suggests that the non-thermal effect may have a significant impact. The same can be said for the inner heliosheath and the ring current. \\

On the other hand, the non-thermal effect of the non-linear perturbed magnetic field due to the ponderomotive force would be relevant for low-temperature plasmas in the same space environments as the temporal term (see table \ref{tab:PF_ParametersTable}), as can be the solar wind and the magnetosheath for what we have that the Washimi and Karpman magnetization can be $23\%$ and $11\%$ larger respectively for the Kappa distribution that in the Maxwellian case. Therefore this non-thermal correction could be relevant to study the perturbation of the ponderomotive force in the interplanetary magnetic field. Indeed, it has been proposed that the ponderomotive force acts as part of a mechanism in the generation of magnetic holes observed in the solar wind \citep{Tsurutani_Dasgupta,Dasgupta2003,Tsurutani2005}. \\

It is worth noting that we are considering frequencies far from the resonances to have an accurate approximation. Nevertheless, due to the above analysis, we know that the non-thermal effect is enhanced with the frequency. Therefore, for larger frequencies, we expect that the non-thermal effects would be even more significant. 

\section{Conclusions}

We have analyzed the consequences of considering the non-thermal effect of magnetized plasmas due to electron and ion cyclotron wave propagation, giving a detailed comparison of the different terms that make up this non-linear force. We have also obtained an expression for the spatial and temporal terms of the ponderomotive force, its magnetic moment, and the nonlinear background magnetic field perturbation for low-temperature plasmas, which can be useful not only for Kappa distributed plasmas if not also for ponderomotive phenomena occurring in a magnetized plasma with finite temperature. 
 
In particular, we have shown that the magnitude of the spatial term of the ponderomotive force is significantly larger for non-thermal magnetized plasmas than for Maxwellian plasmas, having a relative difference of $10^{-1}$ for the electron cyclotron waves with $\omega_{pe}/|\Omega_e| \sim 10^1$ and ion cyclotron waves with $c/c_A \sim 10^4$ and $\beta = 0.1$. Also, for the spatial term of the ponderomotive force, the non-thermal effect increases with the plasma beta for low-temperature plasmas. The same characteristics are found for the temporal factor of the ponderomotive force. We know also that the temporal factor of the ponderomotive force is responsible for the generation of a slowly varying magnetic field in the ponderomotive interaction of the electromagnetic waves with the plasma, which acts as a nonlinear perturbation of the background magnetic field. We have shown that the effect of the kappa distribution in the nonlinear background magnetic field perturbation can be very significant where the induced magnetization can be even  $25\%$ larger than for Maxwellian plasmas for overdense plasmas in our low beta approximation for $\beta_e = 0.1$, $\kappa = 2$ and $\omega/|\Omega_e| = 0.3$. Also, we have shown that the ponderomotive magnetic moment responsible for the MMP force is enhanced for non-thermal plasmas in our low-temperature approximation far from resonances.   
                    
In summary, we have demonstrated that for all terms of the ponderomotive force, the effect of the Kappa distribution characterizing a magnetized plasma can not be neglected. Indeed, even for near cold magnetized plasmas and wave frequency far from the resonance with plasma beta of the order of $\sim 0.1$ and $\omega/\Omega \sim 0.3$ we have shown that the non-thermal effect is very significant.  Hence, our results show that the effect of the kappa parameter must be considered in ponderomotive phenomena related to non-thermal magnetized plasmas which are commonly the characteristics of the plasma in space environments. 

Indeed in section 4, we have also evaluated the effect of the kappa distribution in the ponderomotive force for different space environments, from where we can conclude that is essential to consider the non-thermal effects of the ponderomotive force when studying related phenomena in regions as the solar wind, the magnetosheath, the plasmasheet and the polar cusp, where we can also use our approximation. For other space conditions given for example in the ring current, the inner heliosheath, and the near magnetotail we would have to extend our results for larger values of the plasma beta.  

Also, the analysis given in this research serves as a a solid base from which to extend the study of the non-thermal effects in the ponderomotive force to other modes of propagation for magnetized plasmas as could be waves propagating obliquely to the magnetic field. Besides, it is useful as a starting point from which to include the effect of the Kappa distribution in some ponderomotive phenomena in space physics. The forces analyzed in this work appear in a lot of phenomena that occur due to the interaction of ion cyclotron or electron cyclotron waves with the plasma, as is the case for the acceleration of ions in the polar cusp, auroral density cavities, the penetration of solar wind in the magnetosphere or the electromagnetic ULF waves in the terrestrial magnetosphere \citep{Nekrasov_Feygin_2005,lundin_ponderomotive_2007}. Therefore, the applications of this study can contribute to getting a better understanding of the dynamics of space physics plasmas.  

\begin{acknowledgements}
JE-T acknowledges the support of ANID, Chile through National Doctoral Scholarship No. 21231291. FAA thanks to FONDECYT grant No. 1230094 that partially supported this work. PSM thanks the support of the Research Vice-rectory of the University of Chile (VID) through grant ENL08/23. 
\end{acknowledgements}

%
%

\bibliographystyle{aa}
\bibliography{References}

\begin{appendix}
\section{Dispersion relations}

To obtain the dispersion relation for low-temperature plasmas we use the following asymptotic expression \citep[see equation 55 of ][]{hellberg_generalized_2002}
\begin{eqnarray}
    Z_{\kappa M}(\zeta) \approx -\frac{1}{\zeta}\left(1 + \frac{\kappa}{2\kappa-3}\frac{1}{\zeta^2} + \frac{3\kappa^2}{(2\kappa-5)(2\kappa-3)}\frac{1}{\zeta^4}\right.  \nonumber \\
   \left.+ 15\frac{\kappa^3}{(2\kappa-7)(2\kappa-5)(2\kappa-3)}\frac{1}{\zeta^6}+ \cdots\right).
   \label{eq:Zasympt}
\end{eqnarray}
Using the previous expansion in equation (\ref{eq:varepsilon}) and considering a quasi-neutral plasma composed of electrons and one species of ions we obtain that
\begin{equation}
    \varepsilon_{0\pm}(\omega) = 1 - \frac{\omega_{pe}^2 + \omega_{pi}^2}{(\omega \mp \Omega_i)(\omega \pm \Omega_e)},
\end{equation}
\begin{equation}
    \delta_\pm(\omega) = \frac{1}{2}\left(\frac{\kappa}{\kappa-3/2}\right)\left[\beta_i\frac{\omega \Omega_i^2}{(\omega \pm \Omega_i)^3} + \beta_e\frac{\omega |\Omega_e|^2}{(\omega \mp |\Omega_e|)^3} \right].
\end{equation}
From the previous equations, we can get the dispersion relation for electron cyclotron waves given by equation (\ref{eq:varepsilonColdEC}) by considering $\omega \gg \Omega_i$ and $\omega \gg \omega_{pi}$. Also, we can obtain the ion cyclotron dispersion relation given by equation (\ref{eq:varepsilonColdAW}) by considering that $m_e/m_i \ll 1$. 

\section{Ponderomotive force factors}

Including the dielectric component of the equation (\ref{eq:varepsilon_approx}) in equation (\ref{eq:FPespacial}) we can deduce the factor $f_{(s)}^\kappa$ that accompanies the spatial variation in the ponderomotive force for the Kappa distribution. For both electron waves and ion cyclotron waves we have that
\begin{equation}
    f_{(s)}^\kappa = \frac{1}{16\pi}(\varepsilon_0 - 1) - \frac{1}{16\pi}\varepsilon_0 \delta.
    \label{eq:PFspatialFactor}
\end{equation}.
We have neglected terms of the order of $\delta^2$ and we have considered that $\varepsilon_0 \gg 1$. Using the same procedure we can obtain the temporal factor:
\begin{equation}
\begin{split}
    f_{(t)}^\kappa =& \frac{\varepsilon_0^{1/2} \omega }{16\pi c}\frac{1}{\omega^2}\frac{\partial \omega^2(\varepsilon_0-1)}{\partial \omega} \\
    &- \frac{\varepsilon_0^{1/2}\omega}{16\pi c}\left[\frac{\delta}{2}\frac{1}{\omega^2}\frac{\partial \omega^2 (\varepsilon_0 - 1)}{\partial \omega} + \varepsilon_0 \frac{\partial \delta}{\partial \omega} \right].
\end{split}
\label{eq:PFtemporalFactor}
\end{equation}
Using these equations we can obtain the ponderomotive force factors for electron and ion cyclotron waves.

\end{appendix}

\end{document}